\documentclass[runningheads]{llncs}
\usepackage{graphicx}
\usepackage{graphicx}
\usepackage{color}
\usepackage{caption}
\usepackage{subcaption}
 \usepackage{hyperref} 

\begin{document}
\title{A mirror-Unet architecture for PET/CT lesion segmentation}

\author{Yamila Rotstein Habarnau \and Mauro Nam\'ias }
\authorrunning{Y. Rotstein Habarnau \and M. Nam\'ias}
\institute{Fundaci\'on Centro Diagn\'ostico Nuclear (FCDN), Buenos Aires C1417CVE, Argentina.}
\maketitle              
\begin{abstract}
Automatic lesion detection and segmentation from [${}^{18}$F]FDG PET/CT scans is a challenging task, due to the diversity of shapes, sizes, FDG uptake and location they may present, besides the fact that physiological uptake is also present on healthy tissues.
In this work, we propose a deep learning method aimed at the segmentation of oncologic lesions, based on a combination of two UNet-3D branches. First, one of the network's branches is trained to segment a group of tissues from CT images. The other branch is trained to segment the lesions from PET images, combining on the bottleneck the embedded information of CT branch, already trained. We trained and validated our networks on the AutoPET MICCAI 2023 Challenge dataset. Our code is available at: \url{https://github.com/yrotstein/AutoPET2023_Mv1}.

\end{abstract}

\section{Introduction}

Fluorodeoxyglucose (FDG) positron emission tomography / computed tomography (PET/CT) scans are routinely used for oncologic staging and response assessment. The enormous effort  required by a manual segmentation of all the lesions in a PET/CT scan, makes it infeasible in clinical routine. Counting with automated lesion segmentation methods could therefore represent an important contribution to PET/CT scans analysis and quantification. However, this is still a challenging task, since these are not the only FDG-avid regions, with physiological uptake also present on healthy tissues. Moreover, lesions of different shapes, sizes, and FDG uptake may be found in diverse body regions.  

The limited availability of fully labeled training data represents a  limitation for the development of automated  lesions segmentation methods. In this context, the autoPET-II challenge has been released (as a successor of the autoPET challenge), providing a large publicly available training data set\cite{AutoPETarticle,AutoPET} and aiming at  promoting the research on machine learning-based automated tumor lesion segmentation on whole-body FDG-PET/CT. 

Over the last years, many deep learning methods for the automatic delineation of organs and tissues on CT images have been proposed. This information can serve as anatomic context for PET images. In this work, we make use of two open-source softwares based on artificial intelligence: \textsc{MOOSE}\cite{MOOSE} and \textsc{TotalSegmentator}\cite{TotalSegmentator}. The first one, is capable of delineating 36 anatomical structures in CT images (13 abdominal organs, 20 bone segments, subcutaneous fat, visceral fat, psoas, and skeletal muscle) and 83 brain subregions from PET images. On the other hand, \mbox{\textsc{TotalSegmentator}} segments 104 anatomical structures in CT images (27 organs, 59 bones, 10 muscles, 8 vessels).

We propose in this work a method for automatic segmentation of lesions in PET/CT images based on UNet-3D\cite{UNet,UNet3D} convolutional neural networks. In particular,  we analyse the performance of a variant of the \textit{Mirror} architecture proposed in \cite{Mirror}, which has two UNet-3D branches. First, one of the network's branch is trained to segment a group of tissues from CT image. Then, the other branch is trained to segment the lesions from PET image, receiving on the bottleneck the embedded information of CT branch, already trained. 

We trained our network on the AutoPET MICCAI 2023 Challenge dataset. We evaluated the Dice score of segmented lesions, the false positive volume and the false negative volume on the preliminary test set of the competition.

\section{Materials and Methods}

\subsection{Dataset and preprocessing}

We trained our network on the AutoPET MICCAI 2023 Challenge dataset\cite{AutoPET}, consisting on 1014 whole body FDG-PET/CT studies from 900 patients acquired at the University Hospital Tübingen. 513 of these studies do not show lesions, while the other 501 studies have been diagnosed as malignant melanoma, lymphoma or lung cancer.  We trained our network making use of the following images: PET image in units of standardized uptake values (SUV); CT image resampled to PET resolution; binary mask with 1 indicating the lesion. We only considered in this work the studies with lesions.

Making use of  \textsc{MOOSE} and \textsc{TotalSegmentator} tools, we obtained the tissues segmentation from every CT image. We found that \textsc{TotalSegmentator} had a better performance (specially in abdominal region for images acquired with contrast). We therefore took this as primary segmentation and completed it with those tissues that are only segmented by \textsc{MOOSE}. In this way, we were able to segment 121 tissues.

All the images were cropped to body contour. We performed data augmentation with \mbox{\textsc{TorchIO}}\cite{torchio} library: gaussian blurring, gaussian noise, contrast transformation, rotation, scaling, gamma transformation and mirroring. The range of the parameters and the probabilities for each transformation where set following what is proposed in \cite{nnUNet}. We did not normalize the images, so  voxel values represent SUV in PET image and Hounsfield unit in CT image. 

Finally, we extracted patches of 64$\times$64$\times$64 voxeles and did not keep background patches. In order to balance classes, we considered all the patches with lesions and randomly sampled the same number of patches without lesions (a different sample every epoch).

\subsection{Architecture}

\begin{figure}
    \centering
    \includegraphics[width=0.8\linewidth]{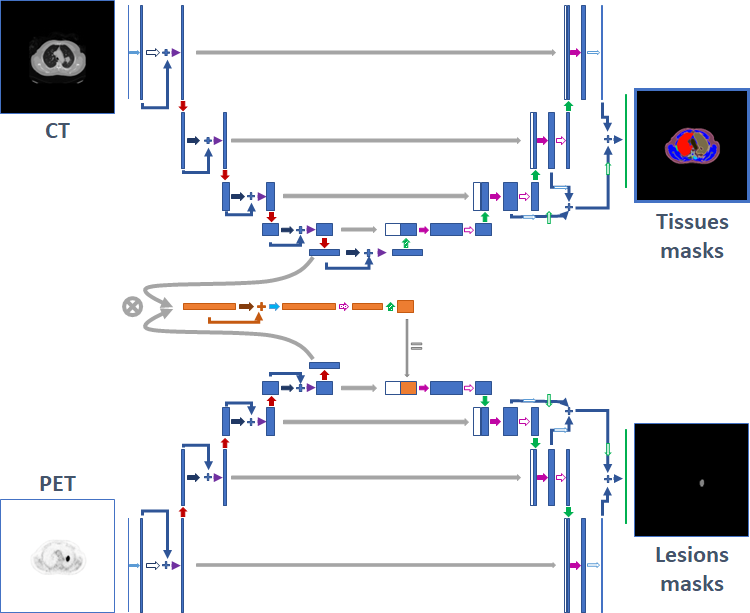}
    \caption{Architecture of our network. The upper branch segments tissues from the CT image, while the lower branch segments the lesions from the PET image, getting the embedded information of CT on the bottleneck.}
    \label{fig:Arq_Mirror}
\end{figure}

We developed a convolutional neural network based on two UNet-3D branches We show in Fig.\ref{fig:Arq_Mirror} a schematic representation of the architecture.  

One of the  branches is trained to segment a group of tissues from CT image. In particular, from the 121 tissues segmented with \textsc{Totalsegmentator} and \textsc{MOOSE}, we got 16 tissues groups: brain, trachea, lungs, adrenal glands, thyroid, spleen, liver, gallbladder, pancreas, urinary system, cardiovascular system, gastrointestinal tract, bones, muscles, fat, others (all the body regions not segmented by either of these tools). These labels where used as reference for the CT branch of the network.

A second branch of the net is trained to segment the lesions from PET image, receiving on the bottleneck the embedded information of CT branch, already trained.

\subsection{Training}

Fist of all, the CT branch was trained for 100 epochs. Then, the weights of this branch were frozen and the PET branch was trained, for 200 epochs.  In both cases, a decaying learning rate was used:
\begin{equation}
    LR(ep) = LR_o \, \left( 1-\frac{ep}{N_{ep}}\right)^{0,9} \, ,
\end{equation}
where $LR_o$ is the initial learning rate, $ep$ the epoch and $N_{ep}$ the total number of epochs trained. For CT branch we fixed $LR_o = 0.01$ while PET branch had  $LR_o = 0.004$. 

Both branches were trained minimizing the sum of BCE and Dice losses. In both cases, the SGD optimizer was used. Besides, in order to improve the  network generalisation, the SWA technique \cite{Izmailov2018AveragingWL} was applied, keeping all the weights every 10 epochs and averaging the last 6 of them (i.e. between epochs 150 and 200).

An 80 GB NVIDIA A100 GPU and a 256 GB RAM CPU were used.

\begin{figure}[ht!]
     \centering
    \begin{subfigure}[b]{0.45\textwidth}
         \centering
         \includegraphics[height=4.4cm]{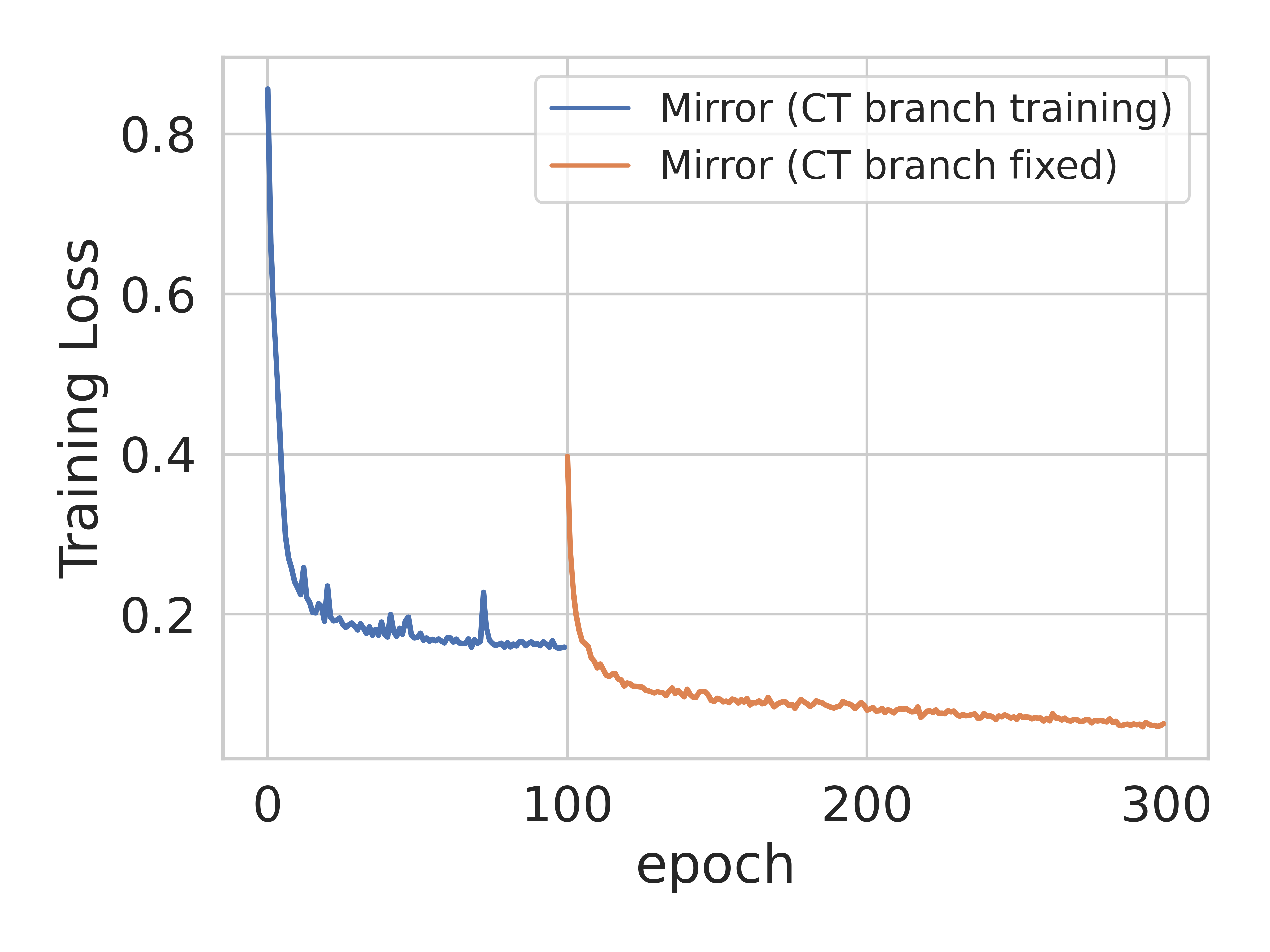}
         \caption{}
     \end{subfigure}
     \begin{subfigure}[b]{0.45\textwidth}
         \centering
         \includegraphics[height=4.4cm]{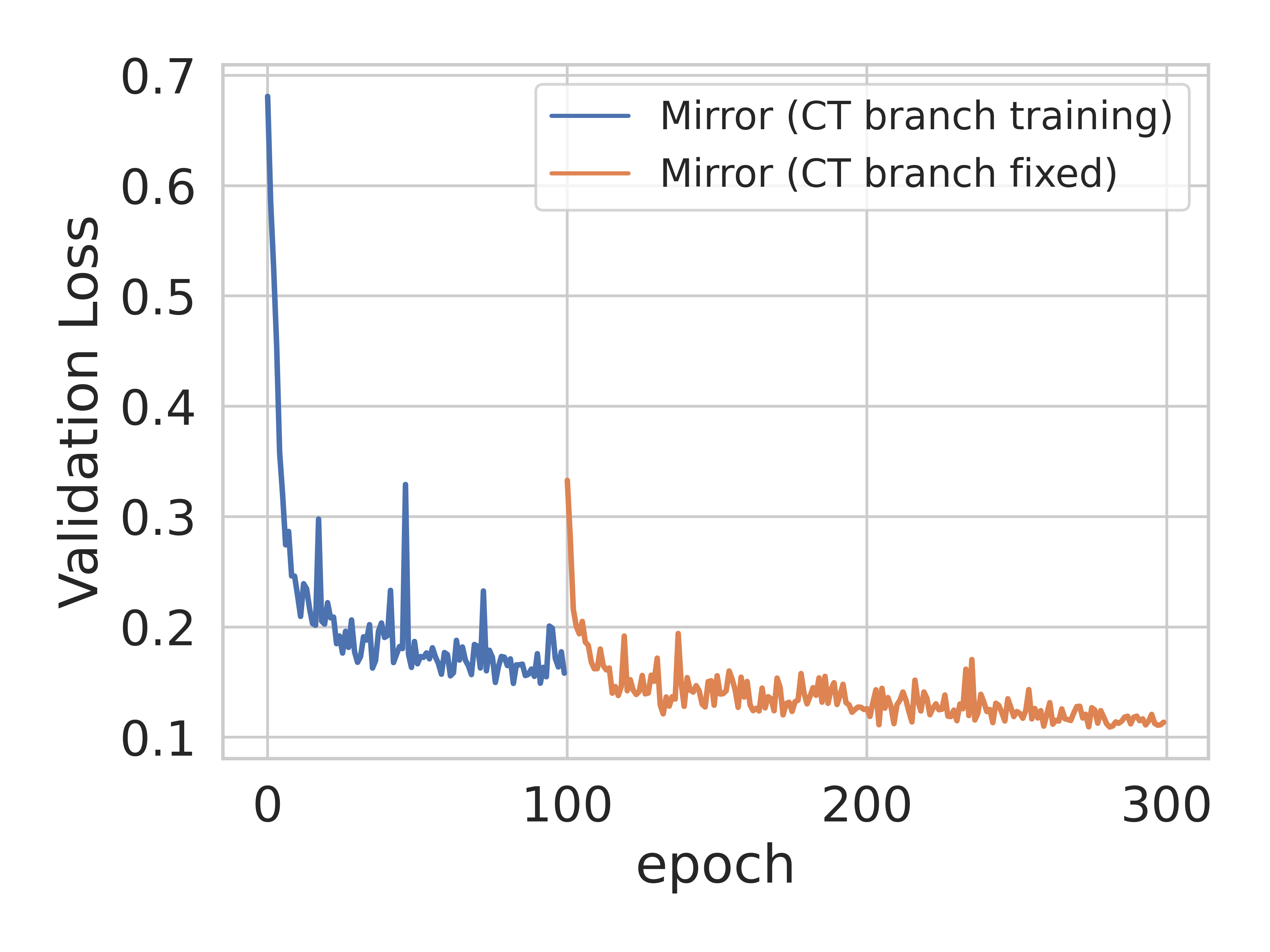}
         \caption{}
     \end{subfigure}
        \caption{Training (a) and validation (b) losses during validation phase (400 training cases and 50 validation cases).}
        \label{fig:Losses}
\end{figure}

As a validation phase, we first split our dataset keeping 400 studies for training, 50 for validation and 51 for test (being careful not to include studies from the same patient in more than one group). We show in Fig. \ref{fig:Losses} the training and validation losses for this phase. 
Finally, we trained over all 501 cases, monitoring the false positive volume on 50 normal cases.

\subsection{Inference and evaluation}

Inference was performed extracting patches of the same size used for training, with adjacent predictions overlapping by half of the size of a patch. The predictions are then combined with  a gaussian importance weight in order to reduce the influence of the voxels from the borders. We applied test time augmentation by averaging the predictions obtained by mirroring along all axes.

The evaluated metrics were:  Dice score, volume of false positive connected components that do not overlap with positives  and volume of positive connected components in the ground truth that do not overlap with the estimated segmentation mask. 

We evaluated our validation phase model on the 50 validation studies. The evaluation of the final model (trained on 501 cases) was performed on the challenge platform on a preliminary  test set, consisting of 5 studies.

\section{Results}

\begin{figure}[ht!]
     \centering
    \begin{subfigure}[b]{0.32\textwidth}
         \centering
         \includegraphics[height=5.6cm]{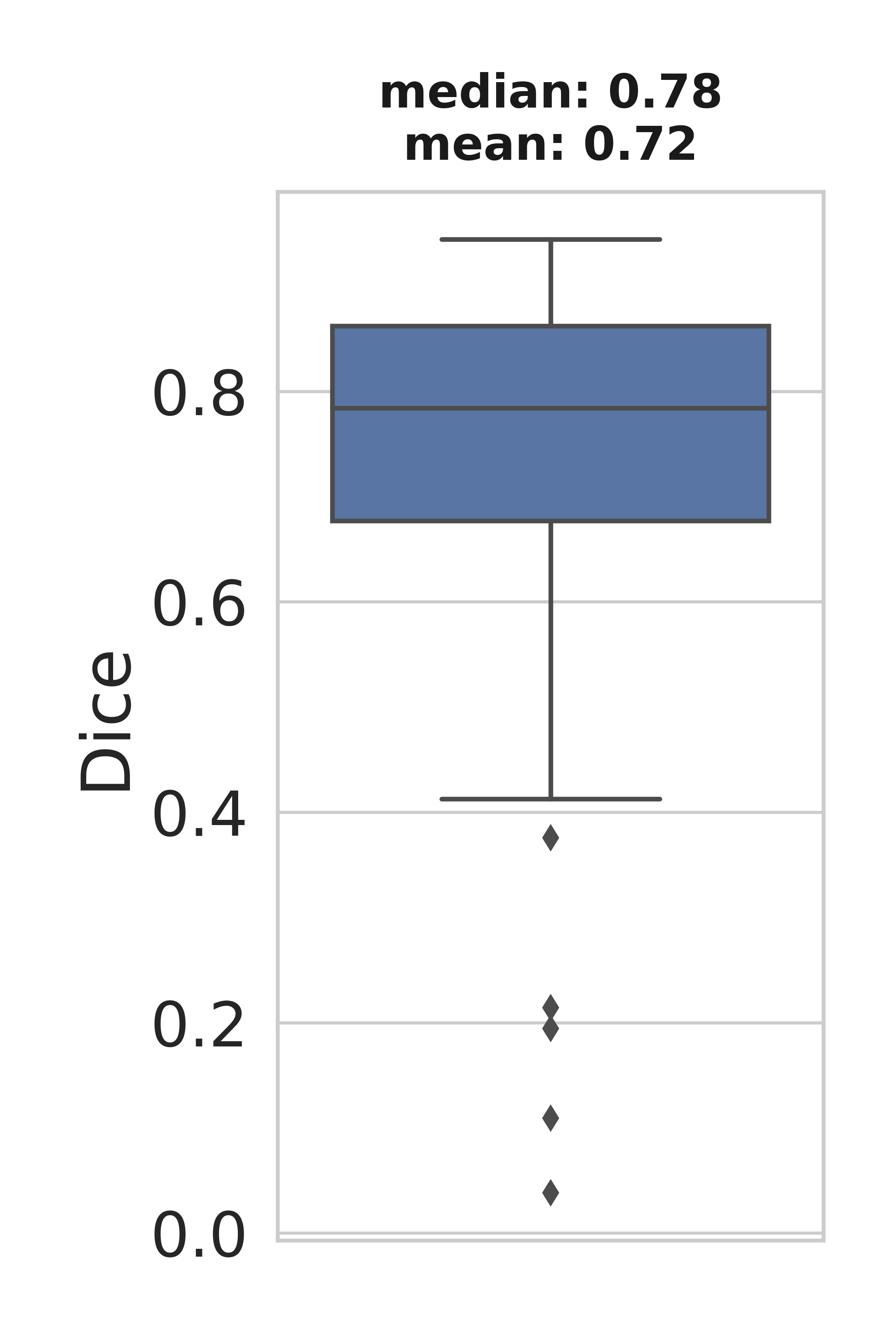}
         \caption{}
     \end{subfigure}
     \begin{subfigure}[b]{0.32\textwidth}
         \centering
         \includegraphics[height=5.6cm]{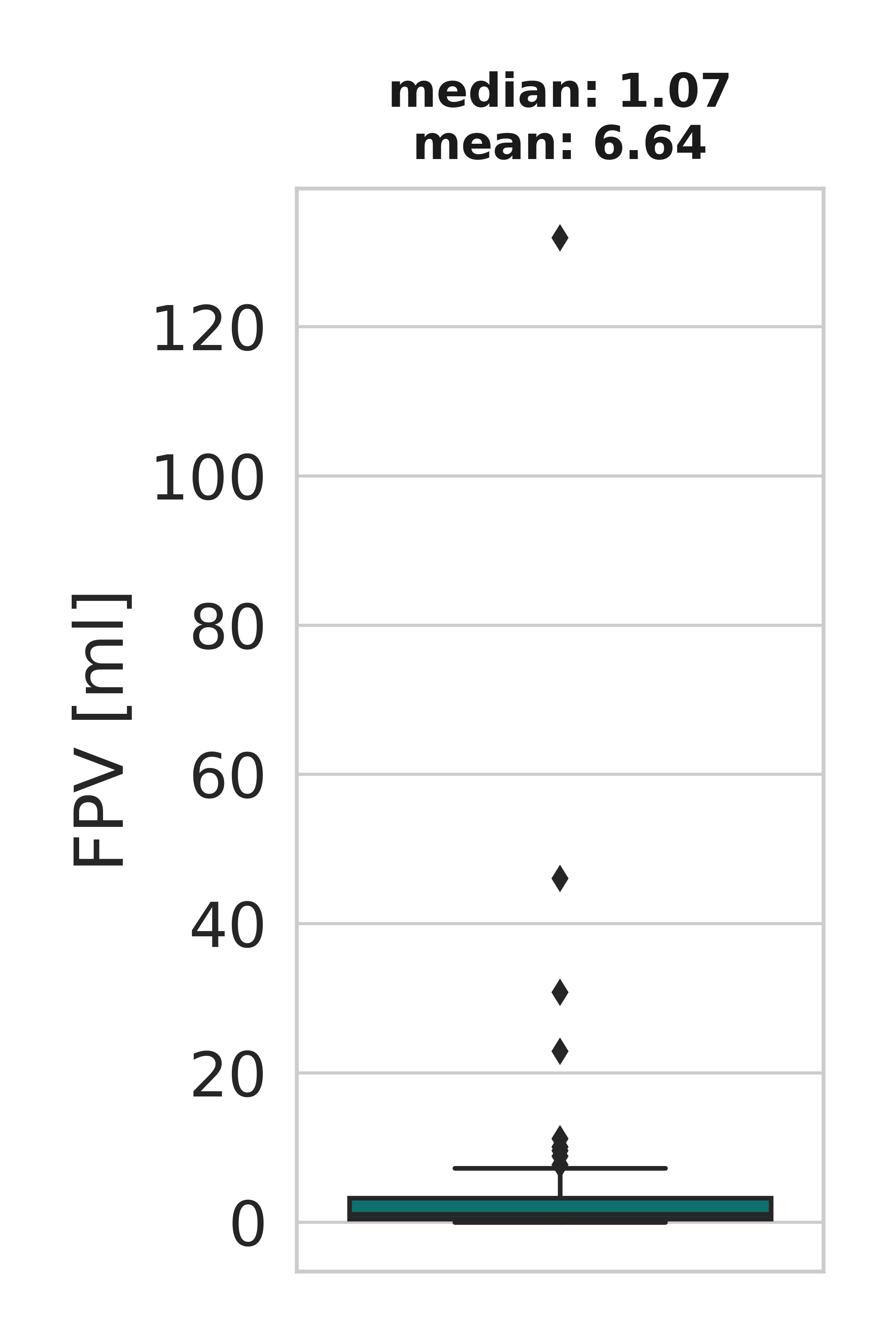}
         \caption{}
     \end{subfigure}
     \begin{subfigure}[b]{0.32\textwidth}
         \centering
         \includegraphics[height=5.6cm]{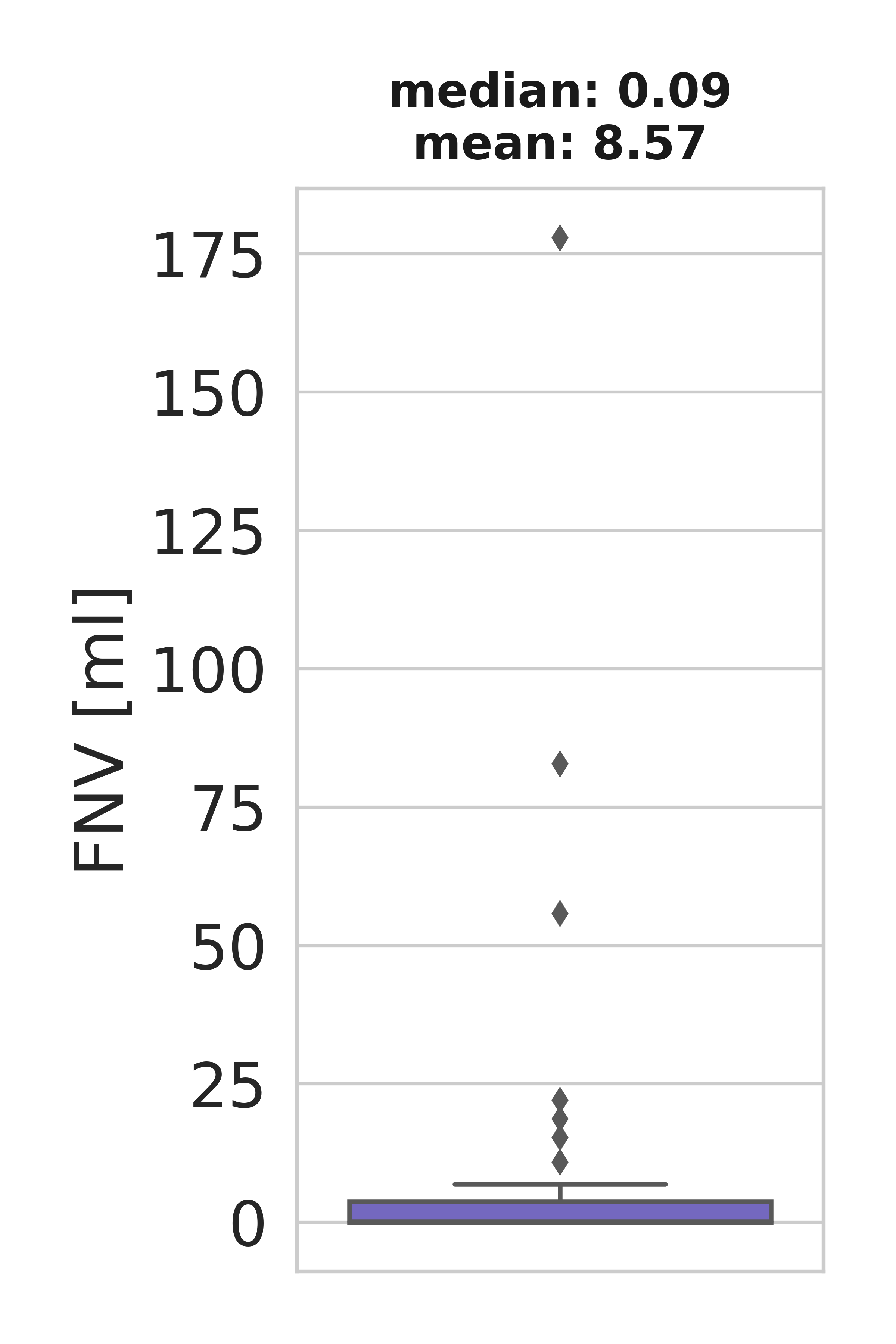}
         \caption{}
     \end{subfigure}
        \caption{Metrics obtained on the 50 validation cases of our validation phase: (a) Dice score, (b) false positive volume, (c) false negative volume.}
        \label{fig:Boxplots}
\end{figure}

We show in Fig. \ref{fig:Boxplots} the distribution of the metrics obtained for the 50 validation cases of our validation phase.

Regarding the final model submitted to the challenge, we got a mean Dice score of 0.54, a mean false negative volume of 0.19 ml and a mean false positive volume of 1.13 ml. We show  in Table  \ref{tab:Metrics} the metrics obtained for each study of the preliminary test set. Is is important to mention that studies with no lesions will have 0 Dice score and 0 false negative volume. 

\begin{table}[ht]
\renewcommand{\arraystretch}{1.5}
  \centering
\begin{tabular}{|c|c|c|c|}
\hline
Case & Dice score &	False negative volume (ml) &	False positive volume (ml) \\ \hline
1 & 0.00 & 0.00 & 0.00 \\ \hline
2 & 0.85 & 0.00 & 0.36 \\ \hline
3 & 0.90 & 0.91 & 2.07 \\ \hline
4 & 0.93 & 0.04 & 1.63 \\ \hline
5 & 0.00 & 0.00 & 1.59 \\ \hline
\end{tabular}
  \caption{Metrics obtained on the preliminary test set.}
  \label{tab:Metrics}
\end{table}

\section{Conclusions}

We developed a deep learning method for lesions segmentation on FDG-PET/CT images, based on two UNet-3D branches. The branch that segments the lesions processes the PET image and receives the embedded information of the CT image on the bottleneck. This embedded information is obtained from the CT branch of the network, that is previously trained for segmenting 16 groups of tissues. 

\bibliographystyle{ieeetr}
\bibliography{bibliografia}

\begin{thebibliography}{10}

\bibitem{AutoPETarticle}
S.~Gatidis, T.~Hepp, M.~Früh, C.~{La Fougère}, K.~Nikolaou, C.~Pfannenberg,
  B.~Schölkopf, T.~Küstner, C.~Cyran, and D.~Rubin, ``A whole-body
  {FDG-PET/CT} dataset with manually annotated tumor lesions,'' {\em Scientific
  Data}, vol.~9, 2022.

\bibitem{AutoPET}
S.~Gatidis and T.~Küstner, ``A whole-body {FDG-PET/CT} dataset with manually
  annotated tumor lesions ({FDG-PET-CT-Lesions}) {[Dataset]},'' 2022.

\bibitem{MOOSE}
L.~K.~S. Sundar, J.~Yu, O.~Muzik, O.~C. Kulterer, B.~Fueger, D.~Kifjak,
  T.~Nakuz, H.~M. Shin, A.~K. Sima, D.~Kitzmantl, R.~D. Badawi, L.~Nardo, S.~R.
  Cherry, B.~A. Spencer, M.~Hacker, and T.~Beyer, ``Fully automated, semantic
  segmentation of whole-body {18F-FDG PET/CT} images based on data-centric
  artificial intelligence,'' {\em Journal of Nuclear Medicine}, vol.~63,
  no.~12, pp.~1941--1948, 2022.

\bibitem{TotalSegmentator}
J.~Wasserthal, H.-C. Breit, M.~T. Meyer, M.~Pradella, D.~Hinck, A.~W. Sauter,
  T.~Heye, D.~T. Boll, J.~Cyriac, S.~Yang, M.~Bach, and M.~Segeroth,
  ``Totalsegmentator: Robust segmentation of 104 anatomic structures in {CT}
  images,'' {\em Radiology: Artificial Intelligence}, vol.~5, no.~5,
  p.~e230024, 2023.

\bibitem{UNet}
O.~Ronneberger, P.~Fischer, and T.~Brox, ``{U-Net}: Convolutional networks for
  biomedical image segmentation,'' in {\em Medical Image Computing and
  Computer-Assisted Intervention -- MICCAI 2015} (N.~Navab, J.~Hornegger, W.~M.
  Wells, and A.~F. Frangi, eds.), (Cham), pp.~234--241, Springer International
  Publishing, 2015.

\bibitem{UNet3D}
{\"O}.~{\c{C}}i{\c{c}}ek, A.~Abdulkadir, S.~S. Lienkamp, T.~Brox, and
  O.~Ronneberger, ``{3D U-Net}: Learning dense volumetric segmentation from
  sparse annotation,'' in {\em Medical Image Computing and Computer-Assisted
  Intervention -- MICCAI 2016} (S.~Ourselin, L.~Joskowicz, M.~R. Sabuncu,
  G.~Unal, and W.~Wells, eds.), (Cham), pp.~424--432, Springer International
  Publishing, 2016.

\bibitem{Mirror}
Z.~Marinov, S.~Reiß, D.~Kersting, J.~Kleesiek, and R.~Stiefelhagen, ``Mirror
  {U-Net}: Marrying multimodal fission with multi-task learning for semantic
  segmentation in medical imaging,'' 2023.

\bibitem{torchio}
F.~Pérez-García, R.~Sparks, and S.~Ourselin, ``Torchio: A python library for
  efficient loading, preprocessing, augmentation and patch-based sampling of
  medical images in deep learning,'' {\em Computer Methods and Programs in
  Biomedicine}, vol.~208, p.~106236, 2021.

\bibitem{nnUNet}
F.~Isensee, P.~F. Jaeger, S.~Kohl, J.~Petersen, and K.~H. Maier{-}Hein,
  ``{nnU-Net}: a self-configuring method for deep learning-based biomedical
  image segmentation,'' {\em Nature Methods}, vol.~18, pp.~203--211, 2021.

\bibitem{Izmailov2018AveragingWL}
P.~Izmailov, D.~Podoprikhin, T.~Garipov, D.~P. Vetrov, and A.~G. Wilson,
  ``Averaging weights leads to wider optima and better generalization,'' in
  {\em Conference on Uncertainty in Artificial Intelligence}, 2018.

\end{thebibliography}

\end{document}